\documentstyle[preprint,prd,aps,psfig,floats,epsf,epsfig]{revtex}
\newcommand{\be}{\begin{equation}}
\newcommand{\ee}{\end{equation}}
\newcommand{\bea}{\begin{eqnarray}}
\newcommand{\eea}{\end{eqnarray}}

\newcounter{dafigcounter}
\tightenlines
\newcommand{\pfig}[3]{
 \refstepcounter{dafigcounter}
 \begin{minipage}[t]{#2}
  \begin{center}
   {\epsfxsize=#2 \mbox{\epsffile{#1.eps}}}
  \end{center}
  \label{#1}
  \small \bf Fig.~\thedafigcounter\rm\ #3
 \end{minipage}
}
\begin{document}
\title{Gluon Pair Production from a Space-Time Dependent Classical 
Chromofield via Vacuum Polarization}
\author{Gouranga~C.~Nayak, Dennis~D.~Dietrich, and Walter~Greiner}
\address{\small\it{Institut f\"ur Theoretische Physik,
J. W. Goethe-Universit\"at,
60054 Frankfurt am Main, Germany}}
\maketitle
%%%%%%%%%%%%%%%%%%%%%%%%%%%%%%%%%%%%%%%%%%%%%%%%%%%%%%%%%%%%%%%%%%%%%%%%%%%%%%%
\begin{abstract}
We investigate the production of gluon pairs from a space-time dependent
classical chromofield via vacuum polarization within the framework of the
background field method of QCD. The investigation of the production of gluon 
pairs is important in the study of the evolution of the quark-gluon plasma in 
ultra-relativistic heavy-ion collisions at RHIC and LHC.
\end{abstract}
\bigskip
%%%%%%%%%%%%%%%%%%%%%%%%%%%%%%%%%%%%%%%%%%%%%%%%%%%%%%%%%%%%%%%%%%%%%%%%%%%%%%%
\section{Introduction}

Ultra-relativistic heavy-ion collisions at RHIC and LHC will provide the best 
opportunity to study the color deconfined state of matter, namely the 
quark-gluon plasma (QGP).
The space-time evolution of the QGP can be split into different stages:
1. the pre-equilibrium,
2. the equilibrium, and 
3. the hadronization stage.
One of the central problems in these experiments is to study how partons are 
formed and how their distribution function evolves in space-time to form an 
equilibrated quark-gluon plasma (if at all).
High momentum partons ($p_T\ge1GeV$), {\it i.e.} minijets are calculated using 
pQCD.
Soft Parton Production is treated differently. There exist various model 
approaches:
1)
In the HIJING model soft parton production is treated via string formations.
2) 
In the color flux-tube model, an extension of the model named before, they 
are treated via the creation of a classical chromofield.
When partons and a classical chromofield are simultanously present, a 
relativistic non-abelian transport equation has got to be solved.

% % % % % % % % % % % % % % % % % % % % % % % % % % % % % % % % % % % % % % % %
\section{Field and Particle Dynamics}

The space-time evolution of the partons can be studied by solving 
relativistic non-abelian transport equations for quarks and gluons 
\cite{nayak,heinz}. 
As the chromofield exchanges color with quarks and gluons, color
is a dynamical quantity. 
The time evolution of the classical color charge 
follows Wong's equations \cite{wong}:

\be
{{{dQ^a} \over d{\tau}}=gf^{abc}u_\mu Q^bA^{c\mu}}.
\ee

There is also a non-abelian version of the Lorentz-force equation:

\be
{{dp^{\mu}} \over d{\tau}}=gQ^aF^{a\mu\nu}u_{\nu}
\ee

Taking the above equations into account, one finds the relativistic 
non-abelian transport equation \cite{nayak,heinz}:

\be
[ p_{\mu} \partial^\mu + g Q^a F_{\mu\nu}^a p^\nu 
\partial^\mu_p
+g f^{abc} Q^a A^b_\mu
p^{\mu} \partial_Q^c ]  f(x,p,Q)=C+S.
\label{trans}
\ee

Note that there are seperate transport equations for quarks, anti-quarks, and
gluons. 
The single-particle distribution function $f(x,p,Q)$ is defined in the 
14-dimensional extended phase space of co-ordinate, momentum, and 
SU(3)-color.
The first term on the LHS of Eq.(\ref{trans}) corresponds to convective flow,
the second to the non-abelian generalization of the effect of the Lorentz 
force, and
the third term describes the precession of the color charge in the presence of 
a classical field.
On the RHS there is the collision term $C$ and the source term for particle
production $S$.
For any system containing field and particles one has the following 
conservation equation:

\be
\partial_\mu T^{\mu\nu}_{mat}+\partial_\mu T^{\mu\nu}_{f}=0
\ee

which is coupled with the above transport equation for the description of the 
QGP. The evolution of the plasma depends crucially on the source term $S$ 
which contains all the information about how partons are produced from the 
classical chromofield.

%%%%%%%%%%%%%%%%%%%%%%%%%%%%%%%%%%%%%%%%%%%%%%%%%%%%%%%%%%%%%%%%%%%%%%%%%%%%%%%
\section{Parton Production from a Space-Time Dependent Chromofield}

The background field method of QCD is a suitable method to describe the 
production of partons from the QCD vacuum via vacuum polarization in the 
presence of a classical chromofield. 
Let us apply the background field method of QCD in order to describe the
production of $q\bar q$-pairs. 

The situation is 
similar to that of $e^+e^-$-pair production described by Schwinger 
\cite{sch51} in QED.
For a space-time dependent classical field $A_{cl}$ 
the amplitude for $e^+e^-$-pair production (see Fig.(\ref{Fig1_})) from the 
vacuum is given by:

\be
M=<k_1,k_2|S^{(1)}|0>~ =~\\
-ie\bar{u}(k_1)\gamma_{\mu}A_{cl}^{\mu}(K=k_1+k_2)
v(k_2)
\ee 

% % % % % % % % % % % % % % % % % % % % % % % % % % % % % % % % % % % % % % % %
\begin{figure}[thb]
\begin{center}
\pfig{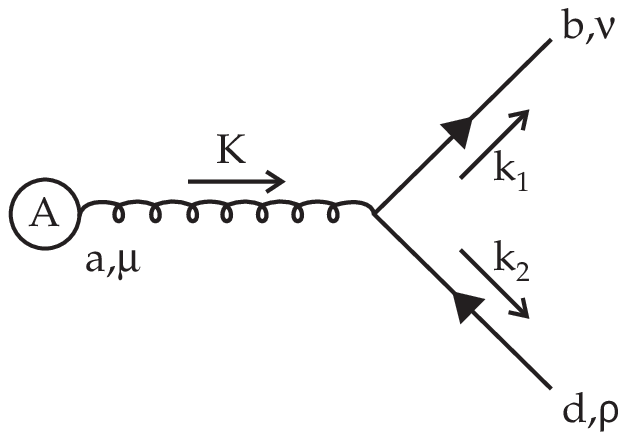}{4.75cm}
{Vacuum polarization diagram for the production of fermions in lowest order}
\end{center}
\end{figure}
% % % % % % % % % % % % % % % % % % % % % % % % % % % % % % % % % % % % % % % %

What, by the general formula:

\be
W^{(1)}=  
\int\frac{d^3k_1}{(2\pi)^3 2k_1^0}\frac{d^3k_2}{{(2\pi)}^3 2k_2^0}
\int d^4K(2\pi)^4\delta^{(4)}(K-k_1-k_2)\sum_{spin}|M|^2
\ee

leads to the pair-production probability \cite{izju}:

\be
W_{e^+e^-}^{(1)}=
\frac{\alpha}{3}\int d^4K ~ 
{(1-\frac{4m_e^2}{K^2})}^{\frac{1}{2}}(1+\frac{2m_e^2}{K^2})
{[|K\cdot A_{cl}(K)|^2-K^2|A_{cl}(K)|^2]}
\ee

where the $d^4K$-integral is defined for $K^2>4m_e^2$.
Simillarly, carrying out the same procedure in the non-abelian theory
one finds for the amplitude for $q\bar{q}$-pair production: 

\be
M=ig\bar{u}^i(k_1)\gamma_{\mu}T^a_{ij}A_{cl}^{a\mu}(k_1+k_2)v^j(k_2)
\ee

and for the corresponding probability \cite{nay99}:

\be
W_{q\bar{q}}^{(1)}=
\sum_{f} \frac{\alpha_s}{6} 
\int d^4K ~
{(1-\frac{4m_f^2}{K^2})}^{\frac{1}{2}}(1+\frac{2m_f^2}{K^2})
~[|K\cdot A_{cl}^a(K)|^2-K^2|A_{cl}^a(K)|^2].
\ee

% % % % % % % % % % % % % % % % % % % % % % % % % % % % % % % % % % % % % % % %
\subsection{Gluon-Pair Production from a Space-Time Dependent Chromofield}

As conventional QCD cannot describe the interaction between a classical 
chromofield and a quantum gluon, one has to fall back on the background field
method of QCD. 
This problem did not arise in QED, as there is no direct interaction between 
the classical field and the photon.
That method was first introduced by DeWitt \cite{dew} and further developped 
by 't Hooft \cite{tho}. In the background field method of QCD, one defines:

\be
A^{a \mu}=
A_{cl}^{a \mu}+A_q^{a \mu},
\ee

where $A_{cl}$ will not be quantized. So the generating functional excluding 
quarks is:

\be
Z[J,A_{cl}]=
\int [dA_q] \det M_G 
\exp(i [S[A_q+A_{cl}]-\frac{1}{2\alpha}G\cdot G + J \cdot A_q]),
\ee

with the classical action:

\be
S[A_q+A_{cl}]=-\frac{1}{4}~\int d^4x ~{(F^{a\mu \nu})}^2,
\ee

where the field-tensor is defined as:

\be
F^{a\mu\nu}=
\partial^{\mu}{(A_q^{a\nu}+A_{cl}^{a\nu})}-
\partial^{\nu}{(A_q^{a\mu}+A_{cl}^{a\mu})}
+g~f^{abc}~{(A_q^{b\mu}+A_{cl}^{b\mu})}{(A_q^{c\nu}+A_{cl}^{c\nu})}.
\ee

The gauge fixing term $G^a$ is chosen  following 't Hooft:

\be
G^a=\partial^{\mu}A_q^{a\mu}+g~f^{abc}~A_{cl}^{b\mu}A_q^{c\mu}.
\ee

The matrix element of $M_G$ is given by:
 
\be
{(M_G(x,y))}^{ab}=\frac{\delta(G^a(x))}{\delta \theta^b (y)}
\ee

which is the functional derivative 
of the gauge fixing term with respect to the infinitesimal change of the 
gauge parameter $\theta$ of the gauge transformation

\be
\delta A_q^{a\mu}= -f^{abc}~\theta^b {(A_q^{c\mu}+A_{cl}^{c\mu})}
~+~\frac{1}{g}~\partial^{\mu}\theta^a.
\ee

Writing $\det M_G$ as functional integral over the ghost field, one obtains 
for the generating functional:

\bea
Z[J,A_{cl},\xi,\xi^{*}]=
\int[dA_q][d\chi][d\chi^{*}]
~\nonumber \\
\times
\exp(i[S[A_q+A_{cl}]+
       S_{ghost}-
       \frac{1}{2\alpha}G\cdot G+J \cdot A_q+\chi^{*} \xi + \xi^{*} \chi]),
\eea

where $\xi$ and $\xi^*$ are source functions for the ghosts and the 
ghost-part of the action is given by:

\bea
S_{ghost}=
-\int d^4x
\chi^{\dagger}_a[\Box^2 \delta^{ab}+
                 g{\overleftarrow{\partial}}_{\mu}f^{abc}(A_{cl}^{c\mu}+
                 A_q^{c\mu})
~\nonumber \\
                 -gf^{abc}A^{c\mu}\partial_{\mu}+
                 g^2f^{ace}f^{edb}A^c_{cl\mu}(A_{cl}^{d\mu}+A_q^{d\mu})]\chi_b.
\eea

Feynman rules involving a classical chromofield, gluons and ghosts
can now be constructed from the above generating functional \cite{abb}.
The vertices involving the coupling of two gluons to the classical field are 
given by:

\be
(V_{1A})^{abd}_{\mu\nu\rho}=
gf^{abd}[-2g_{\mu\rho}K_{\nu}+
g_{\nu\rho}(k_1-k_2)_{\mu}+
2g_{\mu\nu}K_{\rho}]
\ee

for coupling to the classical field once and by

\bea
(V_{2A})^{abcd}_{\mu\nu\lambda\rho}=-ig^2
[f^{abx}f^{xcd}
(g_{\mu\lambda}g_{\nu\rho}-g_{\mu\rho}g_{\nu\lambda}+g_{\mu\nu}g_{\lambda\rho})
~\nonumber \\
+f^{adx}f^{xbc}
(g_{\mu\nu}g_{\lambda\rho}-g_{\mu\lambda}g_{\nu\rho}-g_{\mu\rho}g_{\nu\lambda})
~\nonumber \\
+f^{acx}f^{xbd}
(g_{\mu\nu}g_{\lambda\rho}-g_{\mu\rho}g_{\nu\lambda})]
\eea

for coupling to the classical field twice.

% % % % % % % % % % % % % % % % % % % % % % % % % % % % % % % % % % % % % % % %
\begin{figure}[thb]
\begin{center}
\pfig{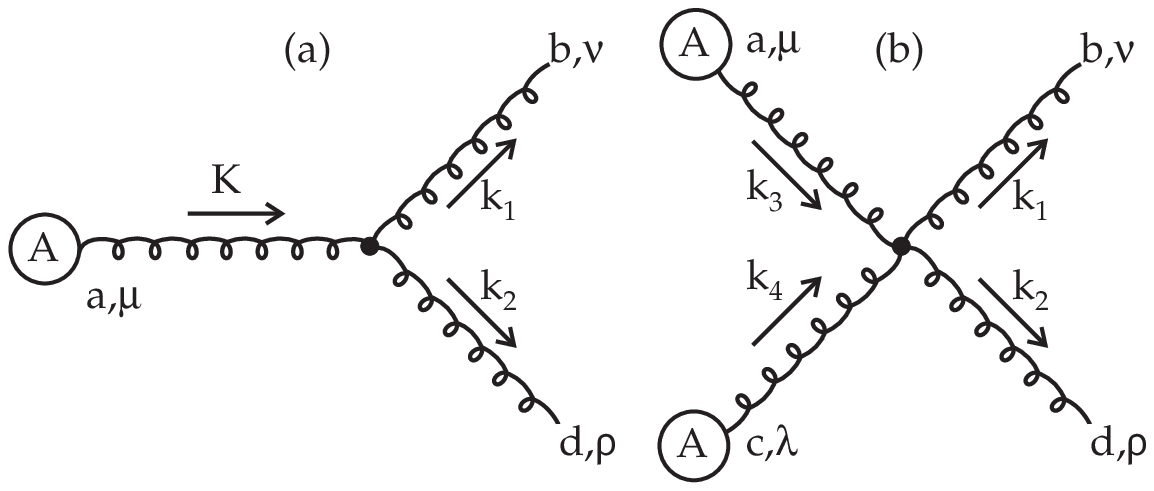}{10cm}
{Vacuum polarization diagrams for the production of gluons in lowest order.}
\end{center}
\end{figure}
% % % % % % % % % % % % % % % % % % % % % % % % % % % % % % % % % % % % % % % %

Note that the above vertices are different from the three and four gluon 
vertices used in conventional QCD and that from hereon the classical field is 
denoted only by $A$ not $A_{cl}$.
The gluon production amplitude $M=<k_1k_2|S^{(1)}|0>$ is defined in a way so 
that $S^{(1)}$ contains all interaction terms of the Lagrangian density 
involving two $Q$-fields, {\it i.e.}:

\newpage

\bea
S^{(1)}=S^{(1)}_G+S^{(1)}_{GF}
~\nonumber \\
=i\int d^4x(
-\frac{1}{2}F^a_{\mu\nu}[A]gf^{abc}Q^{b\mu}Q^{c\nu}
~\nonumber \\
-\frac{1}{2}(\partial_{\mu}Q^a_{\nu}-\partial_{\nu}Q^a_{\mu})
    gf^{abc}(A^{b\mu}Q^{c\nu}+Q^{b\mu}A^{c\nu})
~\nonumber \\
-\frac{1}{4}g^2f^{abc}f^{ab'c'}(A^b_{\mu}Q^c_{\nu}+Q^b_{\mu}A^c_{\nu})
                               (A^{b'\mu}Q^{c'\nu}+Q^{b'\mu}A^{c'\nu})
)
~\nonumber \\
+i \int d^4x(
-\partial_{\lambda}Q^{a\lambda}gf^{abc}A^b_{\kappa}Q^{c\kappa}
~\nonumber \\
-\frac{1}{2}g^2f^{abc}f^{ab'c'}A^b_{\lambda}Q^{c\lambda}
                               A^{b'}_{\kappa}Q^{c'\kappa}
).
\eea

The total amplitude $M=M_{1A}+M_{2A}$ consists of a contribution by the 
three-vertex (see Fig.(\ref{Fig1})(a)):

\bea
M_{1A}=
\frac{(2\pi)^2}{2}\int d^4K\delta^{(4)}(K-k_1-k_2)
~\nonumber \\
A^{a\mu}(K)\epsilon^{b\nu}(k_1)\epsilon^{d\rho}(k_2)(V_{1A})^{abd}_{\mu\nu\rho}
\eea

and one by the four-vertex (see Fig.(\ref{Fig1})(b)):

\bea
M_{2A}=
\frac{1}{4}\int d^4k_3 d^4k_4\delta^{(4)}(k_1+k_2-k_3-k_4)
~\nonumber \\
A^{a\mu}(k_3)A^{c\lambda}(k_4)\epsilon^{b\nu}(k_1)\epsilon^{d\rho}(k_2)
(V_{2A})^{abcd}_{\mu\nu\lambda\rho}.
\eea

The above amplitudes include all the weight factors needed in order 
to retrieve the corresponding Lagrangian density. Now, we again calculate 
the pair production probability:

\be
W=\sum_{spin}
\int\frac{d^3k_1}{(2\pi)^32k_1^0}\frac{d^3k_2}{(2\pi)^32k_2^0}|M|^2.
\ee

To obtain the correct physical gluon polarizations in the final state we use:

\be
\sum_{spin}\epsilon^{\nu}(k_1)\epsilon^{*\nu'}(k_1)=
\sum_{spin}\epsilon^{\nu}(k_2)\epsilon^{*\nu'}(k_2)=-g^{\nu\nu'}
\ee

for the spin-sum and afterwards deduct the corresponding 
ghost contributions.

% % % % % % % % % % % % % % % % % % % % % % % % % % % % % % % % % % % % % % % %
\begin{figure}[thb]
\begin{center}
\pfig{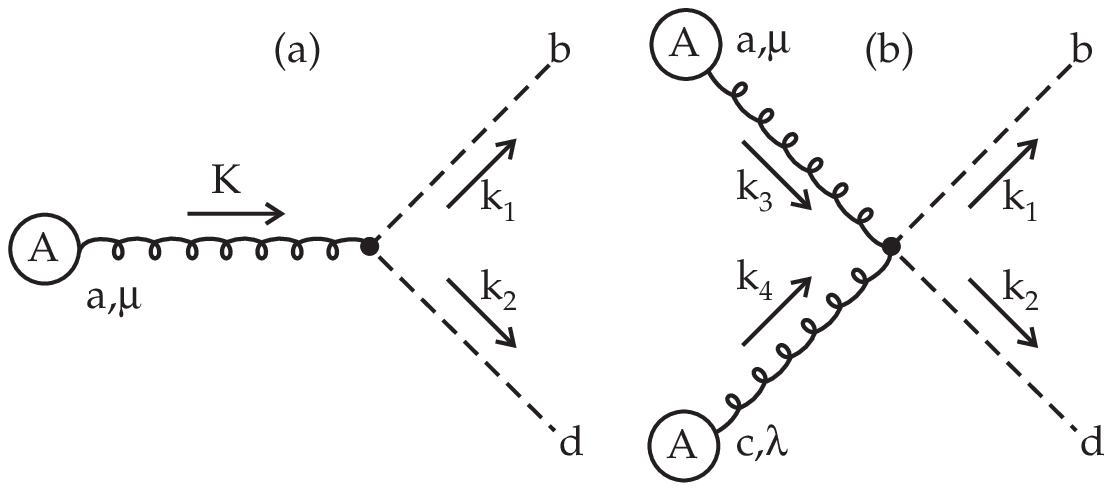}{10cm}
{Vacuum polarization diagram for the production of ghosts in lowest order}
\end{center}
\end{figure}
% % % % % % % % % % % % % % % % % % % % % % % % % % % % % % % % % % % % % % % %

The probability for the gluon part becomes:

\bea
W^{g}
=\frac{10}{8}\alpha_S\int d^4K
~\nonumber \\
((A^a(K)\cdot A^{*a}(K))K^2-(A^a(K)\cdot K)(A^{*a}(K)\cdot K))
~\nonumber \\
+\frac{3ig\alpha_S}{4}\int d^4Kd^4k_3
~\nonumber \\
f^{aa'c'}[(A^a(K)\cdot A^{*a'}(-k_3))(A^{*c'}(K-k_3)\cdot K)]
~\nonumber \\
+\frac{\alpha_Sg^2}{16}\int d^4k_3d^4k'_3d^4K
~\nonumber \\
((A^a(k_3)\cdot A^c(K-k_3))(A^{*a'}(k'_3)\cdot A^{*c'}(K-k'_3))
~\nonumber \\
\times
(f^{abx}f^{xcd}+f^{adx}f^{xcb})(f^{a'bx'}f^{x'c'd}+f^{a'dx'}f^{x'c'b})
~\nonumber \\
+12f^{acx}f^{a'c'x}\times
~\nonumber \\
(A^a(k_3)\cdot A^{*a'}(k'_3))(A^c(K-k_3)\cdot A^{*c'}(K-k'_3))).
\eea

Now, we calculate the ghost part. The vertices involving two ghosts 
and one classical field and two ghosts and two classical fields 
respectively are given by:

\be
(V^{FP}_{1A})^{abd}_{\mu}=+gf^{abd}(k_1-k_2)_{\mu}
\ee

and:

\be
(V^{FP}_{2A})^{abcd}_{\mu\lambda}=
-ig^2g_{\mu\lambda}(f^{abx}f^{xcd}+f^{adx}f^{xcb}).
\ee

The corresponding amplitude for the ghosts reads:

\be
(M^{FP})^{bd}=(M^{FP}_{1A})^{bd}+(M^{FP}_{2A})^{bd}
\ee

with (see Fig.(\ref{Fig2})(a)):

\be
(M_{1A}^{FP})^{bd}=
\frac{(2\pi)^2}{2}\int d^4K(2\pi)^4\delta^{(4)}(k_1+k_2-K)
~\nonumber \\
A^{a\mu}(K)(V^{FP})^{abd}_{\mu}
\ee

and (see Fig.(\ref{Fig2})(b)):

\bea
(M_{2A}^{FP})^{bd}=
\frac{1}{4}\int d^4k_3 d^4k_4\delta^{(4)}(k_1+k_2-k_3-k_4)
~\nonumber \\
A^{a\mu}(k_3)A^{c\lambda}(k_4)(W^{FP})^{abcd}_{\mu\lambda}.
\eea

The probability in this case is simply: 

\be
W^{FP}=
\int\frac{d^3k_1}{(2\pi)^32k_1^0}\frac{d^3k_2}{(2\pi)^32k_2^0}
(M^{FP})^{bd}(M^{FP})^{*bd},
\ee

which becomes:

\bea
W^{FP}=
-\frac{\alpha_S}{8}\int d^4K\delta^{(4)}(K-k_1-k_2)
~\nonumber \\
((A^a(K)\cdot A^a(K))K^2-(A^a(K)\cdot K)(A^a(K)\cdot K))
~\nonumber \\
-\frac{\alpha_Sg^2}{32}\int d^4Kd^4k_3d^4k'_3
~\nonumber \\
(A^a(k_3)\cdot A^c(K-k_3))(A^{a'}(k'_3)\cdot A^{c'}(K-k'_3))\times
~\nonumber \\
(f^{abx}f^{xcd}+f^{adx}f^{xcb})(f^{a'bx'}f^{x'c'd}+f^{a'dx'}f^{x'c'b}).
\eea

The real gluon-pair production probability is given by $W_{gg}=W^{g}-W^{FP}$.

Instead of the probabilities for pair production, one can also consider the 
corresponding source terms which then ultimatively enter the transport 
equation. The source terms are equal to the probability per unit of time and
per unit volume of the phase space. Some calculations yield \cite{ddd}:

\bea
\frac{dW_{q\bar{q}}^{(1)}}{d^4x d^3k}=
\frac{g^2m}{(2\pi)^5\omega}~A^a_{\mu}(x)
~e^{i k \cdot x}
\int d^4x_2 ~A^a_{\nu}(x_2)~e^{-ik\cdot x_2}
~\nonumber \\
(i[k^{\mu} (x-x_2)^{\nu}
  +(x-x_2)^{\mu} k^{\nu}
  +k \cdot (x-x_2)g^{\mu\nu}]
~\nonumber \\ 
(\frac{K_0(m \sqrt{-(x-x_2)^2})m \sqrt{-(x-x_2)^2}+2K_1(m \sqrt{-(x-x_2)^2})}{[\sqrt{-(x-x_2)^2}]^3})
~\nonumber \\
 -m^2 g^{\mu\nu}
 \frac{K_1(m \sqrt{-(x-x_2)^2})}{\sqrt{-(x-x_2)^2}}).
\label{dwqq3}
\eea

for the quarks and:

\bea
\frac{dW_{gg}}{d^4xd^3k}
=
\frac{1}{(2\pi)^5 k^0}\int d^4x' e^{ik\cdot(x-x')}\frac{1}{(x-x')^2}
~\nonumber \\
\times
\{
\frac{3}{4}g^2
A^{a\mu}(x)A^{a\mu'}(x')
[3k_{\mu}k_{\mu'}
-8g_{\mu\mu'}k^{\nu}i\frac{(x-x')_{\nu}}{(x-x')^2}
~\nonumber \\
+5(k_{\mu}i\frac{(x-x')_{\mu'}}{(x-x')^2}
  +k_{\mu'}i\frac{(x-x')_{\mu}}{(x-x')^2})
+\frac{6g_{\mu\mu'}}{(x-x')^2}-12\frac{(x-x')_{\mu}(x-x')_{\mu'}}{(x-x')^4}]
~\nonumber \\
-
3ig^3
A^{a\mu}(x')A^{c\lambda}(x')A^{a'\mu'}(x)
f^{a'ac}K_{\lambda}g_{\mu\mu'}
~\nonumber \\
-
\frac{1}{16}g^4
A^{a\mu}(x)A^{c\lambda}(x)A^{a'\mu'}(x')A^{c'\lambda'}(x')
~\nonumber \\
~[g_{\mu\lambda}g_{\mu'\lambda'}(f^{abx}f^{xcd}+f^{adx}f^{xcb})
                                 (f^{a'bx'}f^{x'c'd}+f^{a'dx'}f^{x'c'b})
~\nonumber \\
 +24g_{\mu\mu'}g_{\lambda\lambda'}f^{acx}f^{a'c'x}]
\}.
\label{dwgg}
\eea

for the gluons.
It can be checked that the above results are gauge invariant with respect to 
type-(I)-gauge transformations \cite{ddd}. 

%  %  %  %  %  %  %  %  %  %  %  %  %  %  %  %  %  %  %  %  %  %  %  %  %  %  %
\section{Discussion}

The above results are still to complicated in order to directly get an idea 
about their content, so we look at them for a special, purely time dependent 
model field.

\be
A^{a3}(t)=A_{in}e^{-|t|/t_0},~t_0>0,~a=1,...,8
\label{as},
\ee

and all other components are equal to zero. 
Many other forms could have been taken. We have 
chosen this option just to get a feeling for how the source term in the 
phase-space behaves.
The actual form of the decay of the classical field can only be 
determined from a self consistent solution of the relativistic non-abelian 
transport equations.
The above choice yields:

\bea
\frac{dW_{q\bar{q}}}{d^4x d^3k}=
16\frac{\alpha_S}{(2\pi)^2}
(A_{in})^2
e^{2 i \omega t}
e^{-|t|/t_0}
\frac{t_0}{1+4\omega^2t_0^2}
\frac{m_T^2}{\omega^2},
\label{dWqqs}
\eea

with $m_T^2=m^2+k_T^2$ where $k_T$ is the transverse momentum and:

\bea
\frac{dW_{gg}}{d^4xd^3k}
=
\frac{24\alpha_S}{(2\pi)^2}(A_{in})^2e^{2ik^0t}e^{-|t|/t_0}
\frac{t_0}{1+4(k^0)^2t_0^2}(-3-\frac{k_T^2}{(k^0)^2})
~\nonumber \\
+
\frac{36\alpha_S^2}{2\pi}(A_{in})^4e^{2ik^0t}e^{-2|t|/t_0}
\frac{t_0}{1+(k^0)^2t_0^2}\frac{1}{(k^0)^2}.
\label{dWtots}
\eea

\begin{figure}
\begin{center}
\refstepcounter{dafigcounter}
\begin{minipage}[h]{5cm}
\epsfig{figure=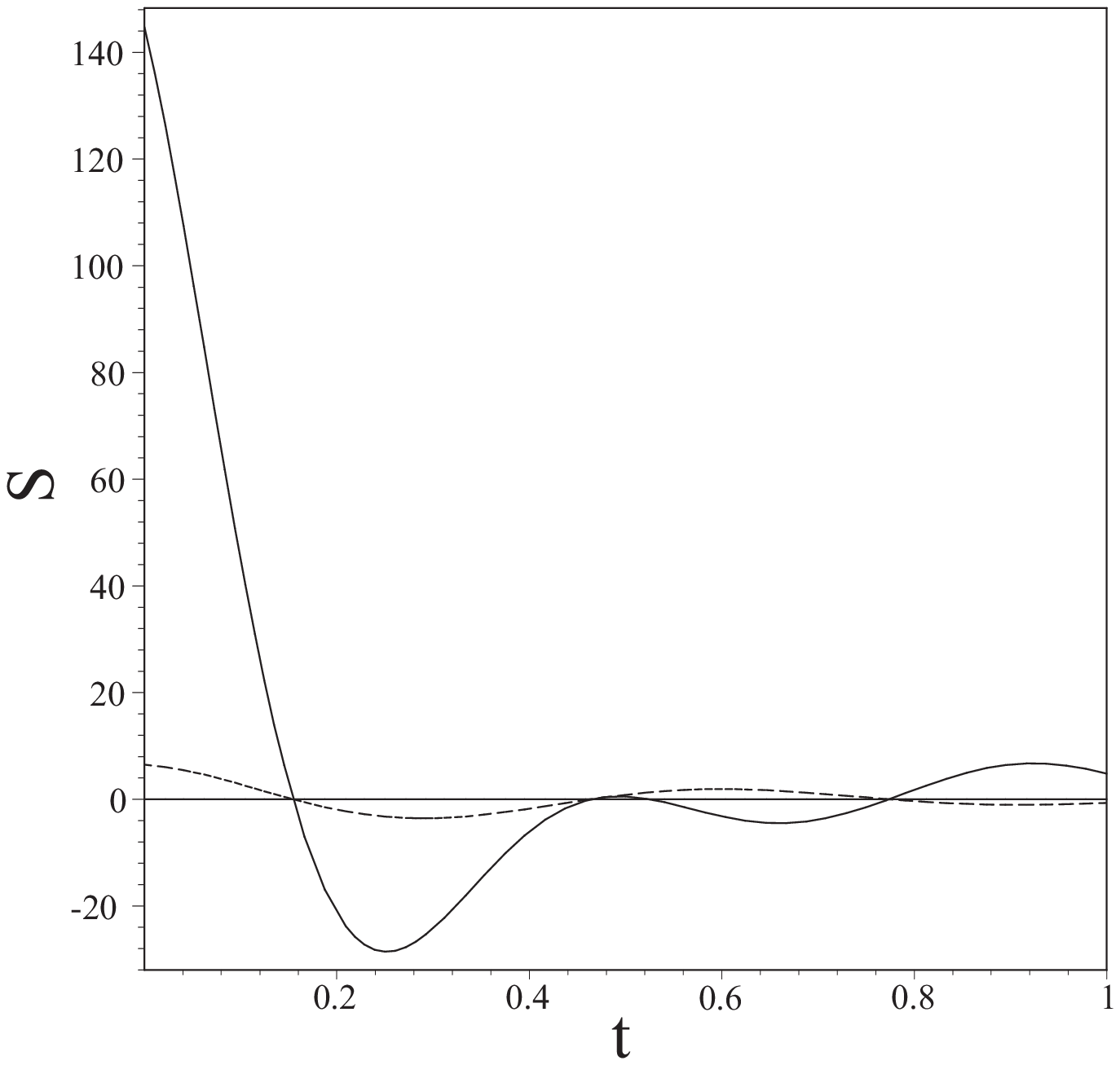, width=4.75cm}
\end{minipage}
\begin{minipage}[h]{5cm}
\epsfig{figure=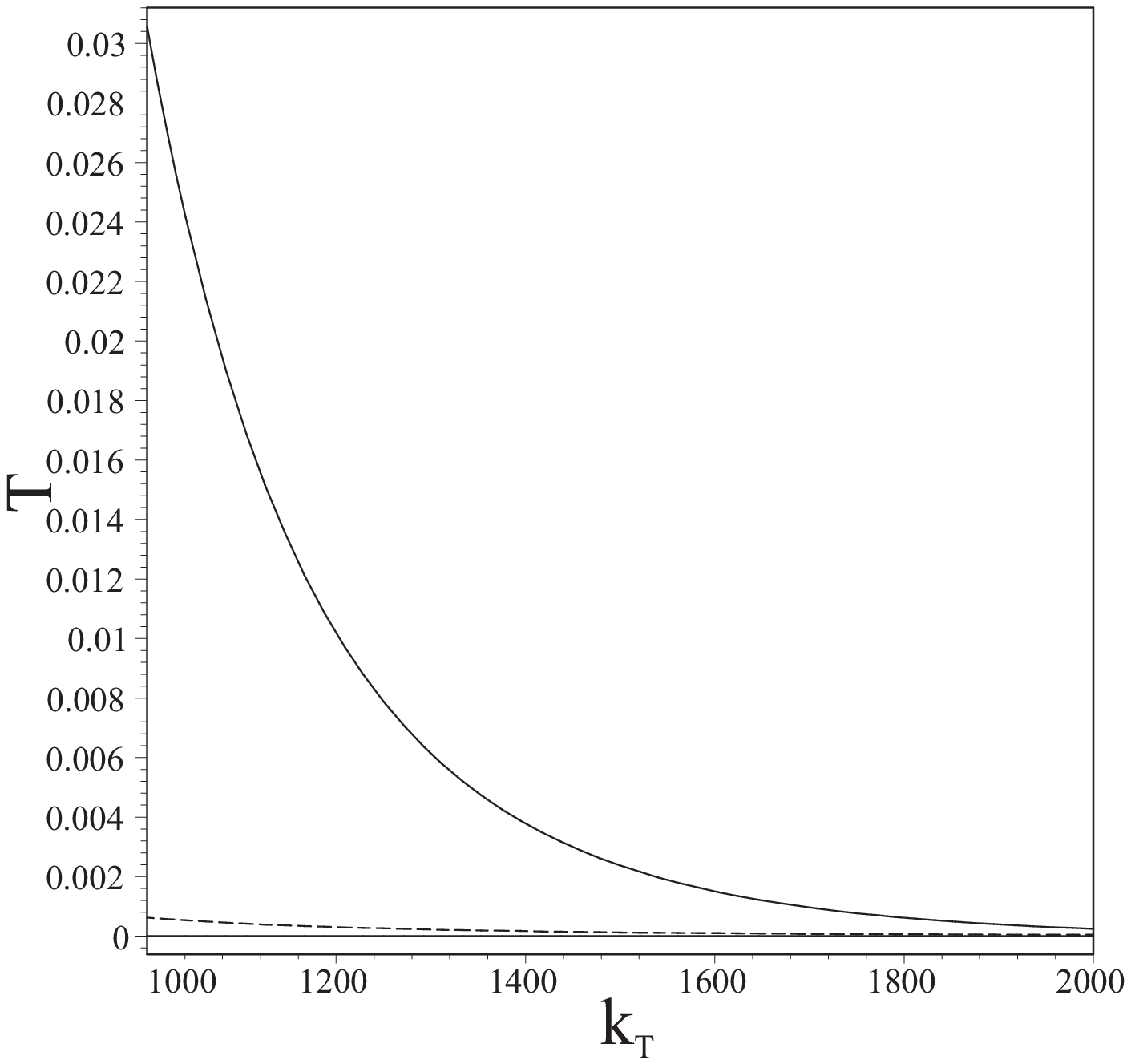, width=5cm}
\end{minipage}
\label{graph}
\end{center} 
\small {\bf Fig.~\thedafigcounter}
\
Source term $S$ [MeV] for quarks (dashed) and gluons (solid) production 
respectively versus 
time $t$ [fm/c] and time-integrated source-tem $T$ for quarks and gluons 
versus $k_T$ [MeV] for the above choice of the model field. 
\end{figure}

We choose the following parameters: $\alpha_S=0.15$, 
$A_{in}=1.5GeV$, $k_T=1.5GeV$, $y=0$, and $t_0=0.5fm$. Additionally, the 
quarks are considered to be massless.
On the LHS of Fig.(\ref{graph}), the oscillatory behavior of the source terms 
$S$ seems to indicate that there exist periods of particle creation 
and particle annihilation which follow each other periodically.
This oscillatory behavior of the source term will play a crucial role once 
it is included in a self consistent transport calculation. It can also be 
seen in the Figure that there are considerably more gluons produced than 
quarks.
On the RHS of Fig.(\ref{graph}), the time-integrated source terms $T$ can be 
regarded as a measure 
for the net-production of particles in an infinitesimal volume around any 
given point in the phase-space. It does not show the oscillatory behavior 
which gives a totally different picture for different times.
In future, we will include these source terms in the transport equation in 
order to study the production and equilibration of the QGP at RHIC and LHC.

% % % % % % % % % % % % % % % % % % % % % % % % % % % % % % % % % % % % % % %  
% % % % % % % % % % % % % % % % % % % % % % % % % % % % % % % % % % % % % % %

%%%%%%%%%%%%%%%%%%%%%%%%%%%%%%%%%%%%%%%%%%%%%%%%%%%%%%%%%%%%%%%%%%%%%%%%%%%%%%%

\end{document}